\documentclass[prl,showpacs,twocolumn,superscriptaddress,floatfix,amsmath]{revtex4}
\usepackage[latin1]{inputenc}
\usepackage[english]{babel}
\usepackage{graphicx}
\usepackage{psfrag}
\usepackage{amssymb}
\usepackage{amsmath}
\usepackage{amscd}
\usepackage{eucal}
\usepackage{color}
\usepackage{bm}
\newcommand{\bi}{\bibitem}
\newcommand{\ignore}[1]{\relax}
\usepackage{epstopdf}
\DeclareMathAlphabet{\mathpzc}{OT1}{pzc}{m}{it}

\newcommand{\e}{{\rm e}}
\newcommand{\rmd}{{\rm d}}
\newcommand{\rmi}{{\rm i}}

\newcommand{\al}{\alpha}
\newcommand{\be}{\beta}
\newcommand{\de}{\delta}

\newcommand{\veps}{\varepsilon}

\newcommand{\tD}{\tau_{\rm D}}

\newcommand{\NL}{N_{\rm L}}

\newcommand{\NR}{N_{\rm R}}
\newcommand{\NS}{N_{\rm S}}
\newcommand{\NSR}{N_{\rm SR}}
\newcommand{\NSL}{N_{\rm SL}}
 
\newcommand{\NT}{N_{\rm T}}

\begin{document}

%%%%%%%%%%%%%%%%%%%%%%%%%%%%%%%%%%%%%%%
\title{
Coherent Thermoelectric Effects in Mesoscopic Andreev Interferometers}
\author{Ph.~Jacquod}
\affiliation{Physics Department,
   University of Arizona, 1118 E. 4$^{\rm th}$ Street, Tucson, AZ 85721, USA}
   \author{Robert S. Whitney}
\affiliation{Institut Laue-Langevin,
6 rue Jules Horowitz, BP 156, 38042 Grenoble, France}
\date{\today}
\begin{abstract}
We investigate thermoelectric transport through Andreev
interferometers. We show that the ratio of the thermal and the charge 
conductance exhibits large oscillations with the phase difference $\phi$ 
between the two superconducting contacts, and that the Wiedemann-Franz law 
holds only when $\phi=\pi$.
A large average thermopower furthermore emerges whenever there is an asymmetry
in the dwell times to reach the superconducting contacts. 
When this is the case, the thermopower is odd
in $\phi$. In contrast, when the average times to reach either superconducting 
contact are the same, the average thermopower is zero, however
mesoscopic effects (analogous to universal conductance fluctuations)
lead to a sample-dependent
thermopower which is systematically even in $\phi$. 
\end{abstract}
% 74.45.+c Proximity effects; Andreev effect; SN and SNS junctions
% 73.23.-b Electronic transport in mesoscopic systems
% 74.78.Na Mesoscopic and nanoscale systems
\maketitle{}
%%%%%%%%%%%%%%%%%%%%%%%%%%%%%%%%%

{\bf Introduction.}
The processing of information unavoidably generates heat~\cite{Lan61}. 
Conventional micro-electronics use electric potentials to switch currents on and 
off. At the nanoscale, however, this becomes energetically 
prohibitive and generates an amount of heat that is hard to dissipate.
As new architectures are explored for quantum communication and 
computing, the question of dissipating heat is again of central 
importance.
In current prototypes for quantum information processors,  
one of the slowest steps is cooling down the qubits
in between  computations~\cite{Bau05},
thus it is crucial to understand heat flows at the 
nanoscale and sub-Kelvin temperatures where quantum coherent effects 
are ubiquitous.

\begin{figure*}[t]
\includegraphics[width=0.95\textwidth]{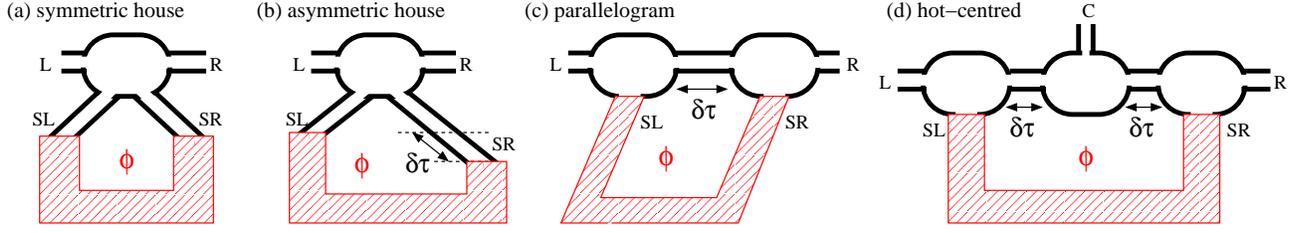}
\caption{\label{fig:models} 
The four Andreev interferometers
considered in this
letter. 
(a)-(b) Single-cavity, two-terminal models, topologically equivalent to the 
house interferometer of Ref.~\cite{Eom98}; the 
symmetry between the leads to the superconducting contacts is broken in model (b).
(c) Double-dot model,
topologically equivalent to the parallelogram interferometer of 
Ref.~\cite{Eom98}.
(d) Triple-dot model topologically equivalent to the hot-middle interferometer of Ref.~\cite{Cad09},
with a hotter lead M.}
\end{figure*}

While quantum interference effects occur in all mesoscopic 
systems~\cite{Imry},
many of them are hugely magnified by Andreev reflection
in normal-metallic/superconducting
nanostructures~\cite{Pet93,Pet95,Har96a,All96}.
Experiments on Andreev interferometers
-- metallic constrictions contacted to two superconducting
terminals with a phase difference,
$\phi$  -- have revealed 
thermoelectric properties that are 
strongly affected by 
these magnified interference effects
~\cite{Veg94,Pet95,Har96a,Eom98,Cad09,Par03,Jia05,Pot94}.
The device properties can be probed (and controlled) 
by varying $\phi$ with an applied magnetic flux or a supercurrent.
For instance, the charge, $G$, and thermal, $\Xi$, 
conductances and the thermopower, $S$, oscillate periodically with $\phi$.
The salient observations are that
(i) the amplitude of the conductance 
oscillations can largely exceeds $e^2/h$ and is typically
larger in samples with larger average conductance, (ii) 
$G$ has its maxima where $\Xi$ has its
minima and vice-versa in violation of the Wiedemann-Franz law, 
(iii) $S$ is significantly larger than in normal metals in absence 
of superconductivity, (iv) $S$  is
either even or odd in $\phi$, depending on the 
interferometer geometry, and (v) $S$ exhibits
oscillations of maximal amplitude at an intermediate temperature.
Despite extensive theoretical investigations~\cite{Bee95,Cla96,Vol96,Naz96,Sev00,Bez03,Vir04,Tit08},
a unified theoretical picture of 
these observations is still lacking. 
In particular, the existing scenarios for an odd
thermopower are always associated with 
a temperature gradient between the two contacts
to the superconducting terminals~\cite{Sev00,Vir04,Tit08}. It is thus
unclear whether these theories can capture 
the recently observed odd thermopower
~\cite{Cad09} in the geometry of Fig.~\ref{fig:models}d, where this 
gradient most likely vanishes.

%%%%%%%%%%%%%%%%%%%%%%%%%%%
%%%%%%%%%%%%%%%%%%%%%%%%%%%

Motivated by these experimental findings, we investigate experimentally relevant
models of Andreev interferometers  
where ideal metallic leads carrying $N_i\gg1$ modes are 
connected at either chaotic ballistic or disordered quantum dots 
with no spatial symmetry.
The dots are contacted to two $s$-wave 
superconductors with order parameters $\Delta e^{i\phi_i}$, each 
carrying ${\NS}_i$ channels,
$i={\rm L,R}$. Physical properties depend
only on the phase difference $\phi_{\rm L}-\phi_{\rm R}$, so we 
set $\phi_{\rm L}=\phi/2$ and $\phi_{\rm R}=-\phi/2$ with $\Delta \in {\cal R}$. We take the
superconductors to be islands through which no current flows on time average in steady-state,
as appropriate to the experiments of Refs.~\cite{Pot94,Pet95,Eom98,Par03,Cad09,Jia05}.
The models are sketched in Fig.~\ref{fig:models}
and are devised to have the same topology as the house
(a and b) and  parallelogram (c) interferometers of Refs.~\cite{Eom98,Jia05}, and (d) 
the interferometer of Ref.~\cite{Cad09} which we call {\it hot-middle}.
These models differ by the absence (a) or presence (b,c,d) of
correlation between the
action phase a quasiparticle accumulates 
and the superconducting phase it acquires at Andreev reflections.
This correlation is key to understanding the large thermopowers induced by
the presence of superconductivity, because it breaks particle-hole
symmetry.

The theory we are about to present gives a unified picture
of thermoelectric transport through mesoscopic Andreev interferometers. Extrapolated to diffusive systems,
it exhibits all the main experimental observations listed above. 
For the asymmetric house (model b), parallelogram and hot-middle interferometers,
it predicts that on average the thermopower will be an odd oscillatory function of $\phi$,
even in the absence of a temperature difference between the two superconducting contacts.
In contrast, for the symmetric-house interferometer it predicts that the average thermopower is zero,
but that mesoscopic fluctuations render the thermopower 
random in sign, but systematically even in $\phi$. 

{\bf Thermoelectric transport.}
The linear response expression for
charge, $I_i$, and heat, $J_i$, currents in lead $i$~\cite{Cla96}
(summation over doubly-occurring indices is assumed) is
\begin{eqnarray}\label{eq:lin_ttrans}
\left(\! \begin{array}{c} I_i \\ J_i \end{array} \!\right)
= -\int_0^\infty \! \! \! \! \rmd \veps F'(\veps)  \left(\! \begin{array}{cc}
\tilde{G}_{ij}(\veps) & \tilde{B}_{ij} (\veps) \\
\tilde{\Gamma}_{ij}(\veps)  & \tilde{\Xi}_{ij}(\veps) 
\end{array} \! \right)
\left( \! \begin{array}{c} V_j\! - \! V_0 \\ T_j\! - \! T \end{array} \! \right) \! ,
\end{eqnarray}
with the derivative $F'(\veps)$ of the Fermi function and the base temperature
$T$.  Since there is no net current into the S loop, its potential, $V_0$, is tuned to ensure  $\sum_j I_j =0$.
In two-terminal geometries, Fig.~\ref{fig:models}a-c, Eq.~(\ref{eq:lin_ttrans})
reduces to
$I = G(V_{\rm L}-V_{\rm R}) + B (T_{\rm L}-T_{\rm R})$
and $J= \Gamma (V_{\rm L}-V_{\rm R}) + \Xi(T_{\rm L}-T_{\rm R})$.

Taking $T_{ij}^{\al \be}(\veps)$ as the transmission coefficient for
a $\beta$-quasiparticle (e or h) injected from lead $j$ 
at energy $\veps$ exiting as an $\al$-quasiparticle in lead $i$,
Ref.~\cite{Cla96} gives
\begin{subequations}\label{eq:th_coeff}
\begin{eqnarray}
\tilde{G}_{ij} (\veps) \! \!&=&\! 
{2e^2 \over h}\big[ 2N_i\de_{ij} 
- T_{ij}^{\rm ee} +T_{ij}^{\rm he} +T_{ij}^{\rm eh} -T_{ij}^{\rm hh}\big],
\\
\tilde{\Xi}_{ij} (\veps)  \! &=&  {2\veps^2 \over hT}
\big[ 2N_i\de_{ij} 
- T_{ij}^{\rm ee} -T_{ij}^{\rm he}- T_{ij}^{\rm eh}-T_{ij}^{\rm hh} \big], \quad \quad
\\ \label{eq:th_coeffB}
\tilde{B}_{ij} (\veps)  \! &=&  -{2e\veps \over hT}
\big[T_{ij}^{\rm ee} -T_{ij}^{\rm he} + T_{ij}^{\rm eh}-T_{ij}^{\rm hh}\big],
\\
\tilde{\Gamma}_{ij} (\veps)  \! &=&  -{2e\veps \over h}
\big[T_{ij}^{\rm ee} +T_{ij}^{\rm he} - T_{ij}^{\rm eh}-T_{ij}^{\rm hh}] \, .
\end{eqnarray}
\end{subequations}

%%%%%%%%%%%%%%%%%%%%%%%%%%%
\begin{figure*}[t]
\includegraphics[width=1.0\textwidth]{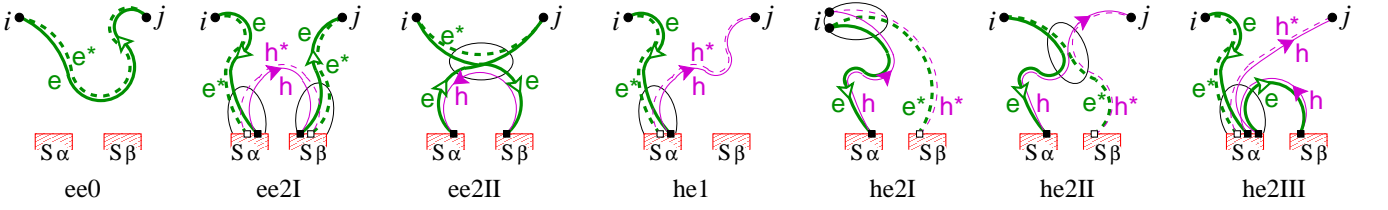}
\caption{\label{fig:T-all} (Color online) Contributions to $\langle T_{ij}^{\rm ee}\rangle$
(first three)
and $\langle T_{ij}^{\rm he} \rangle$ (last four).
Green (violet) paths indicate electrons (holes),
dashed lines indicate complex-conjugated amplitudes and circles show trajectory encounters.
Normal leads are labelled $i,j$ while superconductors are labelled $S\al,S\be$. Contributions to $\langle T_{ij}^{\rm hh}\rangle$ and $\langle T_{ij}^{\rm eh} \rangle$ have e and h interchanged everywhere.
}
\end{figure*}
%%%%%%%%%%%%%%%%%%%%%%%%%%

{\bf Average diagonal thermoelectric coefficients.}
To evaluate the transmission probabilities for Eqs.~(\ref{eq:th_coeff}), we use
the Feynman rules in Ref.~\cite{Whi09},
 taking into account the presence of two S contacts with phase difference $\phi$. 
We  work perturbatively in the ratio $\NS/N$ of the total number of channels 
carried by the S contacts to those carried by the normal leads.  
All contributions up to ${\cal O}[(\NS/N)^2]$ are shown in Fig.~\ref{fig:T-all},
the $\phi$-dependent ones are ee2II, he2I, he2II and he2III. 
For model a and b, their value in Eqs.~(3) and (4) of Ref.~\cite{Whi09} is
now multiplied 
by $g(\phi) = [\NSL^2 + \NSR^2 + 2 \NSL \NSR \cos \phi]/\NS^2$.
For $\NL=\NR=N$, the average
charge and thermal conductances (neglecting weak-localization) 
are
\begin{subequations}\label{eq:conductances}
\begin{eqnarray}
\langle G \rangle  &=& N/2 + \NS^2 \, g(\phi) \, f_G(T) / 8 N \, , \\
\langle \Xi\rangle  & = & \frac{\pi^2 k_{\rm B}^2 T}{3 e^2} (N/2 - \NS^2 \, g(\phi) \, f_\Xi(T) / 8 N) \, .
\end{eqnarray}
\end{subequations}
The main difference between charge and heat conductances in this symmetric configuration is that the backscattering
contribution he2I in $T_{\rm LL}^{he}$ is
absent in $T_{\rm RL}^{he}$, which brings the periodic
oscillations in $G$ and $\Xi$ out of phase by $\pi$, with $G$ being
minimal at $\phi=0$. This fits with the experiment of Ref.~\cite{Jia05}.
The thermal dampings are
polynomial, generalized zeta and polygamma functions of
$\alpha=4 k_{\rm B} T \tau_{\rm D}$, where $\tD$ is the dwell time. 
Asymptotically 
$f_G(0)=1$, $f_\Xi(\alpha \ll 1)\approx 10.4 \alpha$,
and $f_G(\alpha) \approx \pi/\alpha$, $f_\Xi(\alpha)\approx 32/\pi^2 \alpha$, 
for $\alpha \gg 1$.

Eqs.~(\ref{eq:conductances})
imply that there are coherent oscillations of the  Wiedemann-Franz (WF) ratio
\begin{eqnarray}
\frac{\Xi+\Gamma S }{G T} \simeq 
\frac{\Xi}{G T} = l_0 \left( 1-{\NS^2 \, g(\phi) [f_G(T)+f_\Xi(T)] \over  4 N^2} 
\right),
\end{eqnarray}
where our results below show that $\Gamma S$ is small enough to neglect, 
and $l_0=\pi^2 k_{\rm B}^2/3 e^2$ is the Lorenz number. 
Thus, unless $\phi=\pi$ and $\NSL=\NSR$ simultaneously,  
superconductivity causes a 
${\cal O}[(\NS/N)^2]$ violation of the WF law, parametrically
larger than the ${\cal O}[N^{-1}]$ violation induced by mesoscopic
fluctuations in metallic samples~\cite{Vav05}.

{\bf Average thermopower.} The off-diagonal
thermoelectric coefficients satisfy the Onsager relation $B=-\Gamma/T$. We checked
that our theory preserves this symmetry and only discuss $B$ from now on. 
In two-terminal arrangements, $G$ is symmetric in $\phi$. Therefore
the symmetry of the thermopower coefficient $S=-B/G$ is determined by the symmetry of $B$. 
To leading order in $N_i$, particle hole
symmetry, $\veps \rightarrow - \veps$, is equivalent to reversing the superconducting phases,
$\phi \rightarrow - \phi$. Combining this with Eq.~(\ref{eq:th_coeffB}), we 
straightforwardly conclude that $\langle B \rangle$ is generically odd in $\phi$, 
up to weak-localization corrections.
In a house geometry, however, one can interchange the 
superconducting leads, and thus reverse the superconducting phases
without changing the physics. Thus $\langle B \rangle_a$ must be even 
in $\phi$. Neglecting weak localization corrections, one thus has
$\langle B \rangle = 0$ and $S=0$ for model a.

How can a finite leading-order thermopower emerge? 
Our symmetry argument
breaks down when there are correlations between the 
action phase a quasiparticle accumulates on its way through the system
and the superconducting phase that it picks at Andreev reflections.
We find that, when present, 
these correlations generate a finite, odd average thermopower.
This is most easily seen by analyzing the asymmetric house interferometer
of Fig.~\ref{fig:models}b, where a neck renders the journey toward SR 
systematically longer.  
For simplicity, we assume that all trajectories going into
the neck spend a time $\delta \tau$ in it before they 
hit the SR lead. Now take contribution he2I.
If the solid path hits SR and the dashed hits SL,
it induces a phase of $\phi$ from the S leads, and a phase of 
$2\veps \delta\tau$ from the extra length of the solid path.
If the solid path goes to SL and the dashed to SR, 
we get the opposite phases.  Interchanging e and h,
means $\phi \to -\phi$, thus these contributions to 
$T_{ij}^{he}-T_{ij}^{eh}$ behave like 
$\cos(2\veps \delta\tau+\phi )-\cos(2\veps\delta\tau-\phi)$.
The prefactor on this contribution is easily found using the Feynman rules in
Ref.~\cite{Whi09}.
Treating the other contributions in the same way, 
we find that the leading-order average thermopower for model b is
\begin{eqnarray}\label{eq:tp_b}
\langle S_{\rm b} \rangle  \,=\,
-{\langle B_{\rm b}\rangle \over 
\langle G_{\rm b}\rangle} = {4k_{\rm B} \over e} 
{\NSL\NSR\over (N_{\rm L}+N_{\rm R})^2}\  
I_{\rm b} (T) \ \sin \phi, \quad 
\end{eqnarray}
where 
$ I_{\rm b} (T)  =-(k_{\rm B}T)^{-1} \int_0^\infty \rmd \veps  \veps \, F'(\veps)\, \sin(2\veps \de \tau ) /(1+ 4 \veps^2 \tau_{\rm D}^2)$.
This energy integral is zero for $\delta \tau =0$ (symmetric house), is linear in $T$
for $T \ll \delta \tau^{-1}, \tau_{\rm D}^{-1}$, is maximal when
$k_{\rm B}T \sim \delta \tau^{-1} \sim \tau_{\rm D}^{-1}$ 
and decays as $T^{-1}$ when $T \gg  \tau_{\rm D}^{-1}$ for
$\tD \gg \delta \tau$, and as $\exp[-2 \pi k_{\rm B} T \delta \tau]$ for $T \gg \delta \tau$ for
$ \delta \tau \gg \tD$. The average thermopower is always odd in $\phi$, but we stress
that for it to be finite we need a
systematic asymmetry in the distributions of path lengths to
SL and SR. An asymmetry in the probability of hitting the two 
S-contacts, such as for $\NSL\neq \NSR$, is not sufficient.

The presence of the neck with $N_{\rm n}$ channels, in the parallelogram
interferometer of Fig.~\ref{fig:models}c also breaks symmetry between 
the length of paths to SL and paths to SR. We consider $N_{\rm n} \ll N$, and treat the problem to leading order in $N_{\rm n}/N$. 
The two cavities are not symmetric, and might have different dwell times $\tau_{\rm DL}$ and $\tau_{\rm DR}$.  
To leading order in $N_{\rm n}/N$ and $N_S/N$ we obtain
\begin{eqnarray}\label{eq:tp_c}
\langle S_{\rm c} \rangle  = -{2k_{\rm B} \over e} 
{\NSL \NSR N_{\rm n}(\NSL-\NSR)  \, I_{\rm c} (T) 
\, \sin \phi
\over N_{\rm L}N_{\rm R}\big[N_{\rm n}(\NSL+\NSR)+2\NSL \NSR \big]},
\ \ 
\end{eqnarray}
where 
$I_{\rm c} (T)  = -(k_{\rm B}T)^{-1} \int_0^\infty \! \rmd \veps \, \veps \,  F'(\veps)\, 2 {\rm Im} \big[\e^{\rmi 2\veps\de \tau}/A(\veps)\big]$
with $A(\veps)= (1+\rmi 2\veps \tau_{\rm DL})(1+\rmi 2\veps \tau_{\rm DR})$.
Again the thermopower is odd in $\phi$,
but to be finite it requires $\NSL\neq\NSR$. 
In fact, for $\NSL=\NSR$ there are odd-$\phi$ contributions to 
$\langle S_{\rm c} \rangle$ at next order in $N_{{\rm S},n}/N$
(assuming $\tau_{\rm DL} \ne \tau_{\rm DR}$).
We note in passing that a similar expression is obtained for
the hook geometry of Ref.~\cite{Eom98}.

We apply this $(N_{\rm n}/N)$-perturbation theory to the three-dot model of Fig.~\ref{fig:models}d.
The L and R leads have voltages  $V_{\rm L},V_{\rm R}$ such that no current flows in any lead (or S contact) when lead M is held at a temperature $T+\delta T$. 
The thermopower 
$S_{\rm d}^{\rm \al M} = V_{\rm \al}/\de T $, $\al \in {\rm L,R}$, is lead-dependent.
To find this, we solve Eq.~(\ref{eq:lin_ttrans}) for $V_0$,
$V_{\rm L}$ and $V_{\rm R}$ when all charge currents are zero. One obtains
\begin{eqnarray}\label{eq:tp_d}
\langle S_{\rm d}^{\rm LM} \rangle 
=  {2k_{\rm B} \over e}  {N_{\rm nR}^2 N_{\rm nL} \NSL \NSR\, I_{\rm d}(T)  \,\sin \phi \over N_{\rm M} \NL \NR \big[N_{\rm nL}N_{\rm nR} + N'_S(N_{\rm nL}+N_{\rm nR})\big]},\ 
\end{eqnarray}
where $N'_S = 2\NSL \NSR/(\NSL+\NSR)$, and $I_{\rm d}(T)$ is given by
 $I_{\rm c}(T)$ with  $\de\tau \to (\de\tau_{\rm L}+\de\tau_{\rm R})$ and $A(\veps)$ gaining a factor of $(1+2\rmi \veps\tau_{\rm DM})$.
Similarly we find $\langle S_{\rm d}^{\rm RM} \rangle =- (N_{\rm nL}/ N_{\rm nR}) \langle S_{\rm d}^{\rm LM} \rangle$.
Thus for $N_{\rm nR}\simeq N_{\rm nL}$,
one obtains $\langle S_{\rm d}^{\rm RM} \rangle \simeq -\langle S_{\rm d}^{\rm LM} \rangle$.
All this fits with the experimental data or Ref.~\cite{Cad09}.

It is worth recalling that Eqs.~(\ref{eq:tp_b}--\ref{eq:tp_d}) are leading order in 
$\NS/N$, and thus neglect oscillations  
$\propto \NS \cos \phi/N$ in the denominator of $\langle S\rangle$
(coming from $\langle G\rangle$). These terms generate higher 
odd-$\phi$ harmonics, not unlike the experimental findings.

\begin{figure}[t]
\includegraphics[width=5.0cm]{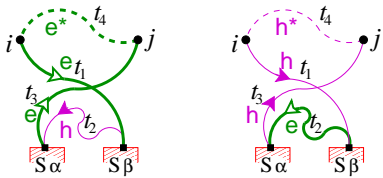}
\caption{\label{fig:no-average} Contributions to ee2II and hh2II before ensemble averaging.}
\end{figure}

{\bf Mesoscopic fluctuations.}
Since the average thermopower vanishes for model a, we look at
mesoscopic fluctuations. We consider contributions
to $(T_{\rm RL}^{ee}-T_{\rm RL}^{hh} + 
T_{\rm RL}^{he}-T_{\rm RL}^{eh})$ before mesoscopic average. The diagrams
shown in Fig.~\ref{fig:no-average} give
a contribution $(T_{\rm RL}^{ee}-T_{\rm RL}^{hh})
\propto \sin[\veps(T-t_2)] \, \sin[\rmi 2E_{\rm F}(T+t_2) + \phi]$
for $T=t_1+t_3-t_4$.
We sum
over the 24 permutations of the four trajectory durations $t_i$, $i=1,2,3,4$, which generally gives
both even-$\phi$ and odd-$\phi$ contributions. For the special case of
the symmetric house, however, the odd contributions cancel out exactly, since
for every contribution touching both superconducting contacts, there is a contribution with equal weight
touching the superconducting contacts in the reversed sequence.
These two come with opposite signs of $\phi$, so the sum is even in $\phi$.
An analysis of $T_{\rm RL}^{he}-T_{\rm RL}^{eh}$ shows the same behavior, therefore
the sample-dependent thermopower must be even in $\phi$ in this case.

To evaluate the typical magnitude $S_{\rm typ}$ of the thermopower 
in a single measurement, we estimate
for ${\rm var} B_a$. This is done by pairing any two contributions
in Fig.~\ref{fig:T-all}, which necessitates to add at least two encounters. From the
Feynman rules in Ref.~\cite{Whi09}, one obtains
the leading order in $\NS/\NT$ by pairing
ee0 with ee0, ee2II and he2III, including all permutations with e$\leftrightarrow$ h.
Then $B_{\rm typ} \approx {\rm rms} [B_a]  \sim (\pi^2 k_{\rm B}^2 T/3 e)
\tD  [ \kappa_1 +\kappa_2 \NSL \NSR \cos(\phi)/N^2 ] $ at low temperature $k_{\rm B} T \tD \ll 1$
where $\kappa_{1,2}={\cal O}[1]$.
The first term is dominated by normal metal fluctuations of $B$~\cite{Lan98}. In contrast to 
Aharonov-Bohm oscillations, mesoscopic fluctuations of thermoelectric coefficients
are not parametrically enhanced by the presence of superconductivity.
A single measurement of the house interferometer of Ref.~\cite{Eom98} thus typically
produces an even$-\phi$ thermopower,
\begin{equation}
S_{\rm typ} \approx ({2k_{\rm B}/e}) \big(S_0 + S_1 \NSL \NSR \cos(\phi)/N^2 \big) ,
\end{equation}
which need not vanish at $\phi=0$.
The constants $S_{0,1}\sim(2e^2/h)G^{-1}\ll1$, so the above $\cos\phi$-term is in principle 
smaller than the odd thermopowers, Eqs.~(\ref{eq:tp_b}--\ref{eq:tp_d}). For a given sample, $S_{0,1}$ are random in sign.
In a symmetric house geometry, 
Ref.~\cite{Eom98} reported an even thermopower at $T=38$mK of similar magnitude as
the odd thermopower found in a parallelogram geometry at $T=350$mK. 
This might be due to 
a much stronger temperature damping in the latter case, or to $N_{\rm SL} \approx N_{\rm SR}$ in the
parallelogram, or both. We note that the experiments 
found $S_{\rm a}(\phi=0) \ne 0$, in agreement with our theory.

%%%%%%%%%%%%%%%%%%%%%%%%%%%
%%%%%%%%%%%%%%%%%%%%%%%%%%%

{\bf Conclusions.} 
Our theory potentially explains all existing thermopower experiments
on Andreev interferometers~\cite{Pet95,Har96a,Eom98,Jia05,Par03,Cad09}.
Finite average thermopowers are systematically odd in $\phi$, and emerge
when the geometry correlates trajectory durations to superconducting contacts
with the phase at the contacts. When the average thermopower vanishes, 
mesoscopic fluctuations cannot be neglected. We discussed the latter for the
first time and showed that they
are systematically even in $\phi$.
Unlike earlier theories~\cite{Sev00,Vir04,Tit08}, our mechanisms for
thermopower do not presuppose
charge imbalance nor require temperature differences 
between the superconducting contacts.
Waiving the latter requirement allows us in particular to explain the 
odd thermopower recently found in hot-middle interferometers~\cite{Cad09}.
We hope that the validity of our theory for even $S$ in house interferometers will soon be checked
by investigations of mesoscopic fluctuations.

This work has been supported by the National Science 
Foundation under Grant No. DMR--0706319. PJ thanks the Theoretical Physics
Department of the University of Geneva and the Aspen Center for Physics for their
hospitality at various stages of this project.
We thank M. B\"uttiker, P. Cadden-Zimansky, V. Chandrasekhar and J. Wei for interesting
discussions.

\end{document}